\documentclass[11pt,a4paper]{article}

\usepackage{jheppub}
\usepackage{booktabs}
\usepackage{placeins}

\graphicspath{{figures/}}

\hypersetup{
pdftitle={Testing bare open-string-metric areas against flavour entanglement in finite-density D3--D7},
pdfauthor={Fuminori Okabayashi}
}

\newcommand{\OSM}{\mathrm{OSM}}
\newcommand{\fl}{\mathrm{fl}}
\newcommand{\ren}{\mathrm{ren}}
\newcommand{\AdS}{\mathrm{AdS}}

\title{
Testing bare open-string-metric areas against\\
flavour entanglement in finite-density D3--D7
}

\author{Fuminori Okabayashi}
\affiliation{Department of Physics, Chuo University,\\
1-13-27 Kasuga, Bunkyo-ku, Tokyo 112-8551, Japan}
\emailAdd{fumi.okabayashi@gmail.com}

\abstract{
We test whether bare codimension-two areas in the D7-brane open string
metric (OSM) reproduce the leading probe-brane contribution to boundary
entanglement entropy in the massless finite-density D3--D7 system at
zero temperature. The established Chang--Karch--Uhlemann benchmark
contains shape-dependent volume terms in a controlled probe regime. By
contrast, the connected strip family terminates at a maximal width and
the selected strip finite part saturates at the disconnected-reference
value, while the lower-area spherical branch is area-like rather than
volume-like at large radius.
The OSM construction nevertheless detects the density-induced crossover
scale, and its geometric landmarks occur at order-one multiples of a
pole-defined longitudinal response length. A low-temperature regulator
shows that no density-controlled horizon-volume term survives the
zero-temperature limit. Bare OSM areas therefore diagnose open-sector
crossover physics but do not, in general, compute flavour entanglement
entropy.
}

\begin{document}
\maketitle
\flushbottom

\section{Introduction}
\label{sec:introduction}

Within gauge/gravity duality
\cite{Maldacena1997,GubserKlebanovPolyakov1998,Witten1998}, probe
branes provide a controlled way to add flavour degrees of freedom to
holographic gauge theories \cite{KarchKatz2002}. At leading order in the
flavour-to-colour ratio \(N_f/N_c\), the
flavour contribution to the boundary entanglement entropy (EE) can be
obtained from the response of the closed-string Ryu--Takayanagi (RT)
area \cite{RyuTakayanagi2006} to the linearised
probe-brane backreaction \cite{ChangKarch2013}, or, in suitable static
Euclidean settings, from generalised gravitational entropy
\cite{LewkowyczMaldacena2013} as adapted to probe branes in
ref.~\cite{KarchUhlemann2014}. An independent direct calculation using a
known first-order backreacted massive D3--D7 geometry is given in
ref.~\cite{KontoudiPolicastro2014}; it is distinct from the Chang--Karch
Green-function construction. In overlapping static examples, the two
methods have been shown to agree at leading probe order, subject to the
corresponding counterterm and boundary-term prescriptions
\cite{ChangKarch2013,KarchUhlemann2014}. The probe-action route avoids
solving explicitly for the backreaction.

The open string metric (OSM) is a physically meaningful effective
geometry for the probe sector and governs the kinetic and causal
structure of probe-brane fluctuations
\cite{SeibergWitten1999,KimShockTarrio2011}. In driven states or at
finite temperature it may develop an effective horizon and temperature
\cite{KimShockTarrio2011,NakamuraOoguri2013,KunduKundu2013,
Kundu2015EffectiveTemperature}. These facts
make it natural to ask whether an RT-type
construction in the OSM can encode entanglement associated with the
flavour degrees of freedom. The standard RT and covariant
Hubeny--Rangamani--Takayanagi (HRT) prescriptions are
formulated for codimension-two surfaces in dynamical closed-string
gravitational geometries
\cite{RyuTakayanagi2006,HubenyRangamaniTakayanagi2007}. The static RT
formula was subsequently placed on a gravitational replica footing,
under the corresponding assumptions, by Lewkowycz and Maldacena
\cite{LewkowyczMaldacena2013}. Applying the same bare area functional to
the non-dynamical OSM is an additional hypothesis.

The practical appeal of a bare-area prescription in the OSM is
particularly strong for an electrically driven probe-brane
non-equilibrium steady state (NESS). Although the Karch--Uhlemann
probe-action method \cite{KarchUhlemann2014} gives direct access to the
leading flavour
contribution to EE in suitable static Euclidean settings, no comparably
direct Euclidean replica implementation is established for a
Lorentzian source such as \(A_x=-Et\). The Chang--Karch
linear-response framework \cite{ChangKarch2013} is formally more
general: worldvolume gauge fields enter through the probe stress
tensor, and Lorentzian
perturbations require retarded bulk response. A steady probe current
does not, however, by itself specify the full stationary state once
backreaction is included. One must also specify how the injected energy
and momentum are absorbed, or equivalently which bath or sink maintains
the steady state.

A closely related antecedent is the closed-string backreaction
calculation of ref.~\cite{OBannonProbstRodgersUhlemann2017}, which
established a first law for entanglement rates in an
electric-field-driven D3--D7 system. In that setting, the
Dirac--Born--Infeld (DBI) worldvolume horizon played no special role in
the entanglement calculation, and the resulting contribution did not
appear to be proportional to the horizon area. The authors also
emphasised the absence of a canonical Newton constant for such a
horizon. Their result constrains the time-dependent entanglement
response, but does not determine the complete flavour EE of a
stationary state with a specified bath.

Against this background,
ref.~\cite{BanerjeeBhattacharyaMaulik2021} applied the standard RT area
functional directly to asymptotically anti-de Sitter (AdS) open-string
geometries in electrically driven probe-brane systems. They considered
strip and spherical subregions, including horizon-induced thermal
volume-law behaviour at large subregion size. The authors described as a
`vital leap of faith' the assumption that the asymptotically AdS
character of the metric is sufficient for applying RT, `irrespective
of the origin' of that metric. The proposal thus applies the RT
functional to a non-dynamical, non-Einstein effective geometry. They
also noted that a generalised-gravitational-entropy analysis would be
needed to establish the prescription unambiguously. Before relying on
such an unweighted area as a surrogate for flavour EE in a driven
state, a minimal consistency test is to examine a static state in which
the leading flavour EE is independently known and bath-dependent
ambiguities are absent.

To our knowledge, the codimension-two bare OSM subregion-area
prescription has not previously been benchmarked quantitatively
against an independently established result for flavour entanglement in
the same state. The static finite-density state studied below is not
intended as a model of a driven NESS. It instead provides a prerequisite
consistency test of the more basic premise that an unweighted OSM area
can serve as a flavour-entanglement functional.

More broadly, holographic entanglement entropy in non-local theories can
exhibit area-to-volume-law crossovers and strong sensitivity to
ultraviolet/infrared (UV/IR) mixing and the geometric regulator
\cite{BarbonFuertes2008,FischlerKunduKundu2013,KarczmarekRabideau2013}.
These works are analogies rather than direct precedents for the
probe-flavour OSM prescription studied here.

The goal of this paper is to perform this static consistency test in
the massless D3--D7 system at finite baryon density and zero
temperature. The leading closed-string RT surface in
the unbackreacted probe geometry does not see the density \(q\), whereas
bare OSM extremal surfaces do see the density-induced scale
\(q^{-1/3}\). The established probe-brane flavour EE contribution in
this system was computed by Chang, Karch and Uhlemann (CKU)
using the leading linearised backreaction
\cite{ChangKarch2013,ChangKarchUhlemann2014}. In the
controlled large-region probe regime, it exhibits \(q\)-dependent volume
terms with non-trivial shape dependence. We compute the corresponding
bare OSM extremal surfaces in the same state and find that their strip
saturation scale and spherical branch endpoints are set by
\(q^{-1/3}\), but their infrared scaling is qualitatively different.

Our conclusion is therefore two-sided: bare OSM minimal areas do not
generally compute the boundary flavour entanglement entropy, but they do
define a useful geometric diagnostic of the density-induced open-sector
crossover scale.
To test the latter interpretation independently, we also compute the
nearest static complex-momentum pole in the longitudinal \(U(1)_B\)
channel and compare its pole-defined response length with the OSM
landmarks.

The paper is organised as follows. Section~\ref{sec:setup} defines the
finite-density OSM and discusses its singular infrared endpoint, while
section~\ref{sec:benchmark} states the probe-brane flavour entanglement
benchmark. Sections~\ref{sec:strip} and \ref{sec:sphere} analyse strip
and spherical OSM surfaces. Section~\ref{sec:comparison} compares their
large-region behaviour with the flavour benchmark,
section~\ref{sec:static-longitudinal} compares the OSM landmarks with an
independent static longitudinal response scale and
section~\ref{sec:discussion} discusses the interpretation and outlook.
Appendices~\ref{app:strip-integrals}--\ref{app:numerics} give
derivations and numerical details; appendix~\ref{app:finite-temperature}
supplies a finite-temperature regularisation of the endpoint and
appendix~\ref{app:probe-flavour-ee} recalls the status of the
complementary probe-brane entanglement methods. Appendix~\ref{app:longitudinal-poles}
derives the static longitudinal pole calculation.

\section{Finite-density D3--D7 and the open string metric}
\label{sec:setup}

We consider the massless D3--D7 system at zero temperature and finite
density \cite{KarchOBannonFiniteDensity2007}. The boundary theory is
\(\mathcal N=4\) \(SU(N_c)\) super-Yang--Mills theory coupled to
\(N_f\) massless \(\mathcal N=2\) hypermultiplets in the fundamental
representation, treated in the quenched probe limit \(N_f\ll N_c\)
\cite{KarchKatz2002}. Here \(N_c\) and \(N_f\) count colours and
flavours, respectively. The conserved radial electric displacement on
the D7-brane is dual to the diagonal flavour \(U(1)_B\) charge density
\cite{KarchOBannonFiniteDensity2007}. We set the AdS radius and
\(2\pi\alpha'\) to one, where \(\alpha'\) is the Regge slope. The
constant \(T_0\) includes the D7-brane tension and the volume of the
internal \(S^3\). The quantity \(q\) denotes this displacement divided
by \(T_0\), and is therefore the reduced charge-density parameter.
Writing
\(\vec x=(x,y,w)\), the induced \(\AdS_5\) part of the D7 worldvolume
metric is
\begin{equation}
ds^2
=
\frac{1}{z^2}
\left(
-dt^2+d\vec x^{\,2}+dz^2
\right).
\end{equation}
The finite-density state is described by a worldvolume gauge field
\begin{equation}
A=A_t(z)\,dt.
\end{equation}
After integrating over the internal \(S^3\), the DBI action is
\begin{equation}
S_{\rm D7}
=
-T_0\int d^4x\,dz\,
\frac{1}{z^5}
\sqrt{1-z^4 A_t'(z)^2}.
\end{equation}
Here \(F_{zt}=A_t'\) is the radial worldvolume field strength, whereas
\(T_0q\) is its conserved conjugate electric displacement. Expressing
the radial field strength in terms of the conserved displacement gives
\begin{equation}
A_t'(z)
=
\frac{qz}{\sqrt{1+q^2z^6}}.
\label{eq:Atprime}
\end{equation}
Anticipating the two classes of boundary regions considered below, we
denote by \(\ell\) the width of an infinite strip, by \(V_2\) its
regulated transverse area, and by \(R\) the radius of a spherical
region. All geometric quantities below are even in \(q\), so we take
\(q>0\). Since \([q]=\mathrm{length}^{-3}\), the density-induced length
scale is \(q^{-1/3}\), and the combinations
\[
q^{1/3}z,\qquad q^{1/3}\ell,\qquad q^{1/3}R
\]
are dimensionless. For a signed density, \(q^{-1/3}\) is replaced by
\(\lvert q\rvert^{-1/3}\).

Writing \(g_{ab}\) for the induced closed-string worldvolume metric and
\(F=dA\) for the worldvolume field strength, the open string metric is
\begin{equation}
G^{\OSM}_{ab}
=
g_{ab}-(Fg^{-1}F)_{ab}.
\end{equation}
For the background above, the static open string metric is
\begin{equation}
ds_{\OSM}^2
=
\frac{1}{z^2}
\left[
-\frac{dt^2}{1+q^2z^6}
+d\vec x^{\,2}
+\frac{dz^2}{1+q^2z^6}
\right].
\label{eq:finite_density_osm}
\end{equation}
This zero-temperature geometry has no regular horizon at finite \(z\).
Its infrared behaviour is controlled by the density-induced crossover
scale \(q^{-1/3}\).

By the bare OSM area we mean the codimension-two area computed from the
five-dimensional metric in eq.~\eqref{eq:finite_density_osm}, without an
additional position-dependent open-string coupling, dilaton factor or
other replica-derived weight. For the massless embedding considered
here, the internal sphere contributes only an overall constant and does
not affect the scaling comparison.

For orientation, the standard static RT prescription considers all
admissible extremal surfaces of codimension two. They are anchored on
the boundary entangling surface and homologous to the chosen boundary
region; the surface with the smallest renormalised area is selected
\cite{RyuTakayanagi2006,LewkowyczMaldacena2013}. Here we imitate only
this variational comparison using the spatial OSM area functional.
Since the OSM is not a dynamical Einstein geometry and no gravitational
replica construction supplies an OSM homology condition, the
lowest-area candidate is selected only within the prescription defined
here. Accordingly, `extremal surface' denotes any solution of the area
Euler--Lagrange equations, whereas `lowest-area candidate' is used
only after the area comparison. This construction does not establish an
entropic interpretation. We therefore treat the bare areas as candidate
open-sector diagnostics and test them below against the established
boundary flavour EE\@.

\subsection{The singular infrared endpoint}
\label{sec:singular-endpoint}

Although the finite-density OSM has no horizon at zero temperature, its
\(z\to\infty\) endpoint is not a smooth cap. The Ricci scalar of
the five-dimensional OSM sector in eq.~\eqref{eq:finite_density_osm} is
\begin{equation}
{\cal R}_{\OSM}
=
-2\,
\frac{
13q^4z^{12}+5q^2z^6+10
}{
1+q^2z^6
},
\end{equation}
and therefore
\begin{equation}
{\cal R}_{\OSM}\sim -26q^2z^6
\qquad
(z\to\infty).
\end{equation}
The singular endpoint lies at finite spatial proper distance: on a
constant-time radial curve,
\(\int_\Lambda^\infty dz/[z\sqrt{1+q^2z^6}]
\sim 1/(3q\Lambda^3)\). By contrast, radial null curves obey
\(dt=\pm dz\), so the endpoint is at infinite static coordinate time.
Its spatial transverse area density
\(\sqrt{G_{xx}G_{yy}G_{ww}}\) equals \(z^{-3}\) and therefore vanishes
at the endpoint.

The endpoint-reaching candidates, introduced explicitly in
sections~\ref{sec:strip} and \ref{sec:sphere}, are the disconnected
strip reference surfaces and the spherical cylinder branch. They extend
to this singular endpoint, but their deep-IR area tails are convergent
and strongly suppressed. The disconnected radial
integrand behaves as \(1/(qz^6)\), while the cylinder integrand behaves
as \(\rho_0^2/(qz^6)\); their contribution above an IR cutoff
\(\Lambda\) is therefore proportional to \(1/(5q\Lambda^5)\). Within
the bare prescription, a formal endcap at the singular endpoint has zero
area because its transverse area density vanishes. The convergence of
these tails makes the bare OSM area functional well defined within the
prescription considered here.

Appendix~\ref{app:finite-temperature} shows that the comparison can also
be formulated using a low-temperature regulator with a regular OSM
horizon.

\section{Benchmark: the CKU probe-brane flavour entanglement result}
\label{sec:benchmark}

The two complementary leading-order probe-brane methods are summarised
in appendix~\ref{app:probe-flavour-ee}. The finite-density benchmark
used below was obtained by Chang, Karch and Uhlemann (CKU) using the
Chang--Karch linearised-backreaction route
\cite{ChangKarch2013,ChangKarchUhlemann2014}.
We denote this first correction, including its physical probe
normalisation, by \(S_{\fl}^{(1)}\).
Within the leading probe approximation, they found that large regions
exhibit volume terms with non-trivial shape dependence. We shall not
need their detailed numerical coefficients. The relevant structure is
\[
\begin{aligned}
S_{\fl}^{(1)}({\rm strip})
&=\cdots+c_{\rm strip}\,qV_2\ell+\cdots,
\\
S_{\fl}^{(1)}({\rm sphere})
&=\cdots+c_{\rm sphere}\,qR^3+\cdots,
\end{aligned}
\]
where \(c_{\rm strip}\) and \(c_{\rm sphere}\) are not related by a
universal entropy density. Their detailed large-region coefficients are
given in ref.~\cite{ChangKarchUhlemann2014}, eqs.~(4.3) and (4.4) in the
published journal version, for the sphere and the strip, respectively; the
corresponding equations are (37) and (38) in arXiv:1406.2705v2, in the
same order. The important point for the present comparison is not the
numerical normalisation of \(c_{\rm strip}\) and \(c_{\rm sphere}\), but
the fact that the leading large-region terms are extensive and
shape-dependent. This shape dependence is more
drastic than the logarithmic enhancement of the area law familiar from
Fermi liquids, and CKU interpreted it as evidence for extensive
entanglement among the flavour degrees of freedom. The overall
normalisation of the reduced density \(q\) is
convention-dependent; the comparison below uses only the large-region
scaling and the shape dependence.

The zero-temperature large-region expansion should not be interpreted
as a controlled statement about the strict \(qL^3\to\infty\) limit,
with \(L=\ell\) for a strip and \(L=R\) for a sphere. At fixed non-zero
dimensionless probe parameter \(t_0\sim N_f/N_c\), this limit lies
outside the controlled regime of the linearised zero-temperature
calculation; the probe-order bookkeeping is reviewed in
appendix~\ref{app:probe-flavour-ee}. The D7 source grows
without bound towards the zero-temperature infrared, so the
backreaction cannot be trusted at \(z\to\infty\). CKU instead restrict
the calculation to large but finite regions for which the backreaction
remains perturbative. For strips, their finite-temperature strip
analysis, with background horizon radius \(z_h\), numerically exhibits
the intermediate hierarchy
\begin{equation}
q^{-1/3}\ll \ell\ll z_h\sim T^{-1},
\label{eq:controlled_probe_window}
\end{equation}
with \(t_0\) also taken sufficiently small that the linearised
backreaction remains perturbative throughout the radial region sampled
by the RT surface. In this regime, the strip volume term persists before
crossing over to the thermal entropy. The window is parametrically
available when
\(\vartheta\equiv\pi Tq^{-1/3}\ll1\). The spherical benchmark used here
is instead the zero-temperature result in a large but finite probe
regime. An analogous finite-temperature hierarchy
\(q^{-1/3}\ll R\ll z_h\) is a parametric expectation, not an explicit
result of the CKU finite-temperature calculation. Here and below, the
CKU large-region result is understood with this distinction.

For the present comparison, the leading \(O(N_c^2)\) adjoint
contribution, computed from the closed-string RT surface in the
unbackreacted probe geometry, is insensitive to the density \(q\),
whereas the leading \(O(N_fN_c)\) probe-brane flavour correction
contains the shape-dependent extensive terms above within the
controlled regimes. The question is whether the bare OSM prescription
reproduces that scaling; sections~\ref{sec:strip}--\ref{sec:comparison}
carry out the comparison.

\section{OSM extremal surfaces for strip regions}
\label{sec:strip}

Consider an infinite strip of width \(\ell\) in the \(x\)-direction,
with transverse area \(V_2\). Two qualitatively different
configurations enter the bare-area comparison. The connected
configuration is a U-shaped extremal surface joining the two components
of the boundary entangling surface and reaching a maximal radial depth
\(z_*\). The disconnected reference consists of two planar sheets
anchored at \(x=\pm\ell/2\), each extending towards the OSM infrared
endpoint. At zero temperature its bare area is independent of
\(\ell\). At fixed \(\ell\), we compare all available connected extrema
with this reference and select the lowest-area candidate within the
prescription defined above. We therefore call this a
disconnected-reference comparison rather than a standard RT phase
transition.

For the connected candidate, on a constant-time slice of
\eqref{eq:finite_density_osm}, write
\begin{equation}
H(z)=1+q^2z^6 .
\end{equation}
The area density is
\begin{equation}
{\cal A}_{\rm strip}
=
V_2
\int dx\,
\frac{1}{z^3}
\sqrt{
1+\frac{z'(x)^2}{H(z)}
}.
\end{equation}
The turning point \(z_*\) is defined by \(z'(x)=0\) at \(z=z_*\).
Since the area Lagrangian has no explicit \(x\)-dependence, the first
integral is
\begin{equation}
\frac{z^{-3}}{\sqrt{1+z'(x)^2/H(z)}}
=
z_*^{-3}.
\end{equation}
Solving this first integral gives
\begin{equation}
z'(x)^2
=
H(z)
\left(
\frac{z_*^6}{z^6}-1
\right),
\end{equation}
and hence
\begin{equation}
\frac{\ell}{2}
=
\int_0^{z_*}
dz\,
\frac{z^3}
{\sqrt{H(z)}\sqrt{z_*^6-z^6}}.
\label{eq:strip_width}
\end{equation}
After setting \(s=z/z_*\) and introducing
\begin{equation}
\zeta=q^{1/3}z_*,
\end{equation}
one obtains
\begin{equation}
q^{1/3}\ell(\zeta)
=
2\zeta
\int_0^1 ds\,
\frac{s^3}
{\sqrt{1+\zeta^6s^6}\sqrt{1-s^6}}.
\end{equation}
Numerically, the connected family reaches a maximal width
\begin{equation}
\boxed{
\ell_{\max}q^{1/3}\simeq 0.7015645.
}
\end{equation}
\begin{figure}[t]
\centering
\includegraphics[width=.78\linewidth]{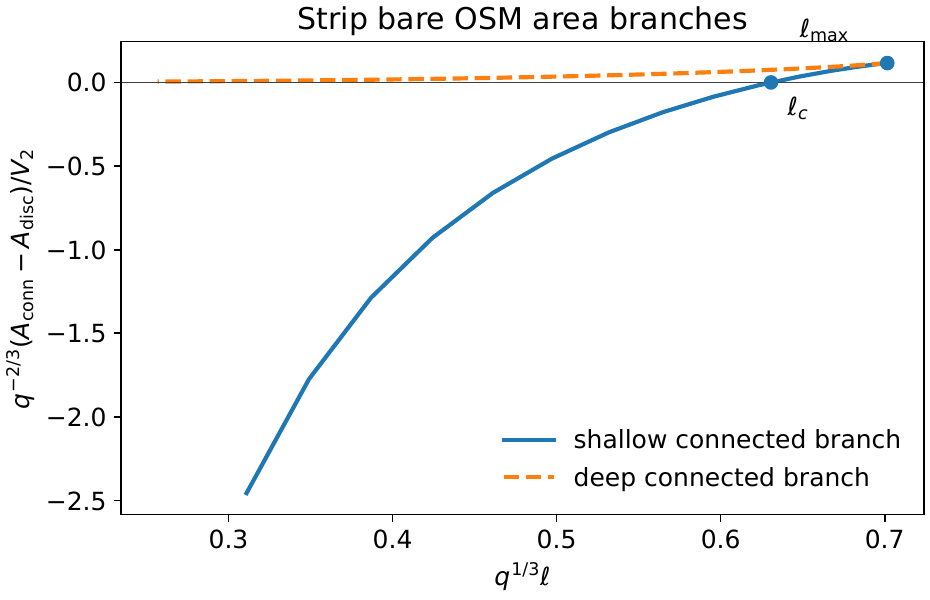}
\caption{
Bare-area difference between the connected and disconnected-reference
configurations for a strip. The vertical axis is the dimensionless
quantity \(q^{-2/3}({\cal A}_{\rm conn}-{\cal A}_{\rm disc})/V_2\), and
the code sets \(q=1\). For any width \(0<\ell<\ell_{\max}\), the
non-monotonic map \(\ell(z_*)\) yields two
connected U-shaped extrema: a shallow branch with smaller \(z_*\) and a
deep branch with larger \(z_*\). They merge at
\(\ell_{\max}q^{1/3}\simeq0.7015645\). The lower-area connected extremum
crosses the disconnected reference at
\(\ell_c q^{1/3}\simeq0.6306986\).
}
\label{fig:strip_branch}
\end{figure}
The function \(\ell(z_*)\) is non-monotonic. Consequently, for a generic
width \(0<\ell<\ell_{\max}\), there are two connected U-shaped extremal
surfaces with the same boundary anchoring: a shallow branch with smaller
\(z_*\), and a deep branch with larger \(z_*\). The two branches merge
at the fold point \(\ell=\ell_{\max}\). These labels distinguish two
extrema of the same connected topology; without a second-variation
analysis, we do not assign either branch a stability label.

Among the two connected extrema, the shallow branch has the lower bare
area. Comparing this lowest-area connected extremum with the
disconnected reference, the area crossing occurs at
\begin{equation}
\boxed{
\ell_c q^{1/3}\simeq 0.6306986.
}
\end{equation}
The connected and disconnected area functionals, their UV-finite
difference, the dimensionless reduction used to determine
\(\ell_c\) and \(\ell_{\max}\) and the quadrature adopted for stable
numerical evaluation are detailed in
appendix~\ref{app:strip-integrals}.
Thus the selected bare OSM area does not produce a large-\(\ell\)
volume law for strips. Instead the connected family terminates at its
fold and the finite part saturates at the value of the disconnected
reference. The connected branches and the area-crossing point are shown
in figure~\ref{fig:strip_branch}.
With a radial ultraviolet cutoff \(z=\epsilon\) and the standard
subtraction, the disconnected value can be written analytically as
\begin{equation}
\frac{{\cal A}_{\rm disc}^{\ren}}{V_2}
=
2\left[
\int_\epsilon^\infty
\frac{dz}{z^3\sqrt{1+q^2z^6}}
-
\frac{1}{2\epsilon^2}
\right]
=
\frac{1}{3}B\left(-\frac13,\frac56\right)q^{2/3}.
\end{equation}
Here \(B(a,b)\) is the Euler beta function. Numerically, the coefficient
is approximately \(-0.86237\).
This convergence and the vanishing endpoint area density follow from
the estimates in section~\ref{sec:singular-endpoint}.

\section{OSM extremal surfaces for spherical regions}
\label{sec:sphere}

Now consider a spherical boundary region
\begin{equation}
\rho^2=x^2+y^2+w^2\le R^2.
\end{equation}
Denoting the constant-time open-string spatial slice by \(\Sigma\), its
metric is
\begin{equation}
ds^2_{\Sigma,\OSM}
=
\frac{1}{z^2}
\left(
d\rho^2+\rho^2d\Omega_2^2
+
\frac{dz^2}{1+q^2z^6}
\right).
\end{equation}
Here \(d\Omega_2^2\) is the metric on the unit two-sphere.
For spherical boundary regions, the extremal-surface equation admits
two qualitatively different infrared behaviours. A cap-type surface
closes off smoothly on the symmetry axis, \(\rho=0\), at a finite radial
position \(z=z_*\). A cylinder-type surface instead approaches a
non-zero radius \(\rho_0\) while extending towards the OSM infrared
endpoint \(z\to\infty\). These names describe the global shape and
infrared behaviour of the extrema, not their area ordering. When
several extrema coexist at fixed \(R\), their renormalised bare areas
are compared before the lowest-area candidate is selected.
The full extremal-surface equations, the regular cap and cylinder
shooting expansions and the critical shrinking separatrix are derived
in appendix~\ref{app:spherical-eoms}.

\subsection{Cap branch}

For cap-like surfaces \(z=z(\rho)\), the area functional is
\begin{equation}
{\cal A}_{\rm cap}
=
4\pi\int_0^R d\rho\,
\frac{\rho^2}{z^3}
\sqrt{
1+\frac{z'(\rho)^2}{1+q^2z^6}
}.
\label{eq:cap_area}
\end{equation}
Regularity at the centre gives
\begin{equation}
z(\rho)
=
z_*
-
\frac{1+q^2z_*^6}{2z_*}\rho^2
+
O(\rho^4).
\end{equation}
Numerically, the cap branch reaches a maximal radius
\begin{equation}
\boxed{
R_{\max}^{\rm cap}q^{1/3}\simeq 0.9948745.
}
\end{equation}

\subsection{Cylinder branch}

A second branch is described by \(\rho=\rho(z)\), with area
\begin{equation}
{\cal A}_{\rm cyl}
=
4\pi\int dz\,
\frac{\rho(z)^2}{z^3}
\sqrt{
\rho'(z)^2+\frac{1}{1+q^2z^6}
}.
\label{eq:cylinder_area}
\end{equation}
The infrared expansion is
\begin{equation}
\rho(z)
=
\rho_0
+\frac{1}{10\rho_0 q^2 z^4}
-\frac{1}{1000\rho_0^3q^4z^8}
+O(z^{-10}).
\end{equation}
The cylinder limit \(\rho_0\to0\) and the cap limit \(z_*\to\infty\)
approach the same unique critical shrinking solution,
\begin{equation}
\rho_{\rm crit}(z)
=
\frac{1}{\sqrt{3}\,qz^2}
+O(z^{-8}).
\label{eq:critical_shrinking_leading}
\end{equation}
This solution is the separatrix between the cap and finite-\(\rho_0\)
cylinder families, rather than an additional continuous family. Direct
shooting from its infrared expansion gives
\begin{equation}
R_\infty q^{1/3}=0.9917717.
\end{equation}
Numerically, the cylinder branch exists for
\begin{equation}
\boxed{
R\ge R_{\min}^{\rm cyl},
\qquad
R_{\min}^{\rm cyl}q^{1/3}\simeq0.9859970.
}
\end{equation}

\begin{figure}[t]
\centering
\includegraphics[width=.78\linewidth]{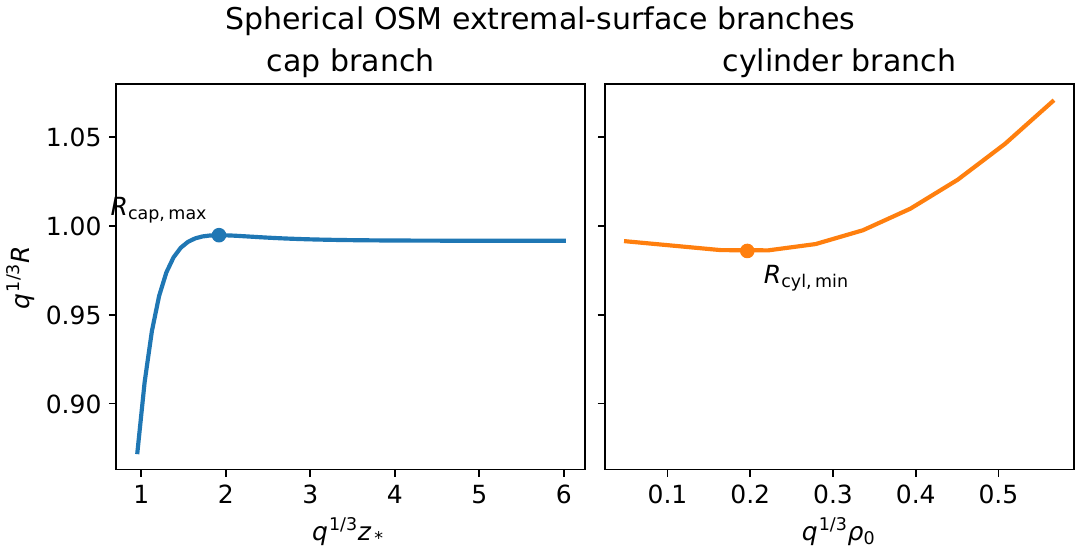}
\caption{
Spherical OSM extremal-surface branches. The cap branch reaches a maximal
boundary radius, while the cylinder branch has a minimal boundary
radius. This figure shows extremal-surface branches, not an area
dominance comparison. The two panels use their natural branch
parameters, \(q^{1/3}z_*\) for caps and \(q^{1/3}\rho_0\) for
cylinders. The overlap region in boundary radius is narrow but
non-zero. The large-\(z_*\) cap and small-\(\rho_0\) cylinder ends meet
at the common limiting radius quoted in the text. The bare-area crossing
in the overlap window is quoted
separately in table~\ref{tab:numerical_values}.
}
\label{fig:sphere_branches}
\end{figure}

Thus the spherical OSM extremal-surface branches show the structure
\begin{equation}
\begin{aligned}
R&<R_{\min}^{\rm cyl}
&&:\quad \text{cap branch only},
\\
R_{\min}^{\rm cyl}&\le R\le R_{\max}^{\rm cap}
&&:\quad \text{one or more cap- and cylinder-type branches coexist},
\\
R&>R_{\max}^{\rm cap}
&&:\quad \text{cylinder branch only}.
\end{aligned}
\end{equation}
The two branch maps are shown in figure~\ref{fig:sphere_branches}.
This statement concerns the existence of branches. In the narrow
overlap window, the area ordering is determined by comparing
\({\cal A}_{\rm cap}^{\ren}(R)-{\cal A}_{\rm cyl}^{\ren}(R)\) at fixed
boundary radius. Since the same boundary-local subtraction is used at
the same \(R\), it cancels in this difference. Numerically, the
lowest-area cap-type and cylinder-type branches cross at
\begin{equation}
\boxed{
R_\times q^{1/3}=0.9897(1).
}
\end{equation}
The parenthetical digit denotes a deterministic numerical envelope, not
a statistical error. Appendix~\ref{app:numerics} details the shooting,
branch tracking, common-cutoff fixed-\(R\) area comparison and construction
of this envelope.
Table~\ref{tab:numerical_values} collects the strip landmarks, the
spherical branch endpoints and the fixed-\(R\) crossing.
The cap-type surface has lower bare area for
\(R_{\min}^{\rm cyl}<R<R_\times\), while the cylinder-type surface has
lower bare area for \(R_\times<R<R_{\max}^{\rm cap}\). We shall not refer
to the cap/cylinder overlap itself as a phase transition.

At large \(R\), the cylinder branch gives
\begin{equation}
{\cal A}_{\rm cyl}^{\ren}
\sim
C_{\rm cyl}\,q^{2/3}R^2+\cdots,
\end{equation}
i.e.\ an area-like scaling rather than a volume law.
The leading coefficient follows from setting
\(\rho(z)=R+O(R^{-1})\) in eq.~\eqref{eq:cylinder_area}:
\begin{equation}
C_{\rm cyl}
=
\frac{2\pi}{3}B\left(-\frac13,\frac56\right)
\simeq
-5.41843.
\end{equation}
We define the displayed finite coefficients by minimal subtraction. In
this scheme, the leading large-radius
cylinder finite part scales as \(R^2\). Since no replica construction
fixes a renormalisation prescription for the bare OSM diagnostic, we do
not assign scheme-independent significance to \(C_{\rm cyl}\) itself.
Nevertheless, admissible boundary-local changes of scheme remain
area-like and cannot generate the \(R^3\) volume term relevant to the CKU
comparison. Thus the robust distinction is between area-like and
volume-like large-region behaviour. The same reasoning leaves the strip
saturation, rather than an \(\ell\)-dependent volume term, intact.

\FloatBarrier

\begin{table}[t]
\centering
\begin{tabular}{lll}
\toprule
Quantity & Branch parameter & Dimensionless value \\
\midrule
\(\ell_{\max}q^{1/3}\) & \(q^{1/3}z_*=1.0208508\) & \(0.7015645\) \\
\(\ell_cq^{1/3}\) & \(q^{1/3}z_*=0.7736443\) & \(0.6306986\) \\
\(R_{\max}^{\rm cap}q^{1/3}\) & \(q^{1/3}z_*=1.92039\) & \(0.9948745\) \\
\(R_{\min}^{\rm cyl}q^{1/3}\) & \(q^{1/3}\rho_0=0.19639\) & \(0.9859970\) \\
\(R_\infty q^{1/3}\) & critical shrinking solution & \(0.9917717\) \\
\(R_\times q^{1/3}\) & fixed-\(R\) cap/cylinder comparison & \(0.9897(1)\) \\
\bottomrule
\end{tabular}
\caption{
Numerical branch endpoints and bare-area crossings for the OSM
surfaces. The calculations use the dimensionless convention \(q=1\);
the displayed combinations restore the density scale. The sphere
crossing compares the lowest-area cap-type and cylinder-type branches
at fixed boundary radius and is the only entry whose numerical
systematics affect the displayed digits.
}
\label{tab:numerical_values}
\end{table}

\section{Large-region comparison}
\label{sec:comparison}

As summarised in table~\ref{tab:large-region-comparison}, the
probe-brane flavour EE contribution and the bare OSM
area involve the same density-induced crossover scale
\(q^{-1/3}\), but have different density-controlled large-region
scaling. The CKU probe-brane result contains shape-dependent
extensive contributions that cannot be reduced to a universal
entropy-density term and were interpreted as evidence for long-range
flavour entanglement, while the bare OSM area measures a different
geometric property of the open-string sector. The finite-density state
therefore provides a counterexample to the general identification of
bare OSM minimal areas with boundary flavour EE\@.

\begin{table}[t]
\centering
{\small
\begin{tabular}{@{}lll@{}}
\toprule
Quantity & Strip & Sphere \\
\midrule
Probe-brane flavour EE & shape-dependent volume term & shape-dependent volume term \\
Bare OSM area & disconnected-reference saturation & area-like cylinder branch \\
\bottomrule
\end{tabular}
}
\caption{
Large-region comparison between the established leading probe-brane
flavour EE and the bare OSM area. The strip flavour-EE entry refers to
the controlled finite-temperature window of
eq.~\eqref{eq:controlled_probe_window}, while the spherical entry refers
to the large but finite zero-temperature probe regime.
}
\label{tab:large-region-comparison}
\end{table}

\section{Static longitudinal response scale}
\label{sec:static-longitudinal}

The mismatch established in section~\ref{sec:comparison} rules out the
bare OSM area as a general flavour-entanglement functional, but leaves
open a narrower question: whether its geometric landmarks nevertheless
track a physical length scale in the probe sector. Since the OSM governs
worldvolume fluctuations, this question can be tested against a response
scale defined independently of the area functional. We use the static
longitudinal \(U(1)_B\) channel, for which \(q\) is the background charge
density and the normalisable zero-frequency problem selects discrete
complex-momentum poles.

Complex-momentum poles of source-free bulk fluctuations determine
pole-defined spatial response scales
\cite{AmadoHoyosLandsteinerMontero2007}. For a purely imaginary pole
\(k=i\kappa\), the corresponding pole contribution carries an
exponential factor \(e^{-\kappa |x|}\), so \(\kappa^{-1}\) defines a
pole-defined response length. We denote the inverse imaginary part of
the nearest discrete longitudinal pole by \(\xi_\parallel^{\rm pole}\). At
zero temperature, this pole-defined scale need not exhaust possible
non-pole or power-law contributions to the full position-space
correlator.

Let \(\mathcal E\) denote the gauge-invariant longitudinal
worldvolume-gauge-field fluctuation. With
\[
u=q^{1/3}z,
\qquad
k=i\kappa_\parallel,
\qquad
c_\parallel=\frac{\kappa_\parallel}{q^{1/3}},
\]
the zero-frequency gauge-invariant D3--D7 fluctuation obeys
\begin{equation}
\frac{d}{du}
\left[
\frac{(1+u^6)^{3/2}}{u}
\frac{d\mathcal E}{du}
\right]
+
c_\parallel^2
\frac{\sqrt{1+u^6}}{u}
\mathcal E
=0 .
\label{eq:longitudinal-static-pole}
\end{equation}
This equation follows from the full quadratic DBI fluctuation action:
its radial weights encode not only the OSM causal structure but also
the DBI measure and effective gauge coupling. The comparison below is
therefore between bare geometric landmarks and an open-sector response
observable, not a claim that the metric alone determines the pole.
This is the zero-frequency D3--D7 specialisation of the finite-density
longitudinal fluctuation equation of
ref.~\cite{KarchSonStarinets2009}. Their momentum \(q_{\rm KSS}\) and
density \(d\) are denoted here by \(k\) and \(q\), respectively. The
source-free ultraviolet condition is \(\mathcal E\sim u^2\), while
infrared normalisability uniquely selects
\(\mathcal E\sim u^{-7}\). The latter condition is also obtained by
introducing the low-temperature regulator of
appendix~\ref{app:finite-temperature} and then taking its
zero-temperature limit.

The lowest eigenvalue is
\begin{equation}
\boxed{
k_{\star,\parallel}
=
i\,3.6334359\,q^{1/3}
},
\qquad
\xi_\parallel^{\rm pole}
=
\frac{1}{\lvert\operatorname{Im}k_{\star,\parallel}\rvert}
=
0.2752216\,q^{-1/3}.
\label{eq:longitudinal-screening-result}
\end{equation}
Here \(\xi_\parallel^{\rm pole}\) is the pole-defined static response
length in the longitudinal density channel of the global \(U(1)_B\)
current.\footnote{If the boundary \(U(1)_B\) is
dynamically gauged, the corresponding screened electrostatic response
is instead determined by a Coulomb-dressed propagator or, equivalently,
mixed boundary conditions.} Combining this result with the geometric
landmarks gives
\begin{equation}
\kappa_\parallel\ell_c=2.29160,
\qquad
\kappa_\parallel\ell_{\max}=2.54909,
\qquad
\kappa_\parallel R_\times=3.5960(4),
\qquad
\kappa_\parallel R_\infty=3.60354 .
\label{eq:geometric-screening-products}
\end{equation}
The products \(\kappa_\parallel R_{\min}^{\rm cyl}\) and
\(\kappa_\parallel R_{\max}^{\rm cap}\) fall in the same narrow range
as \(\kappa_\parallel R_\times\) and \(\kappa_\parallel R_\infty\).
This clustering mainly reflects the narrow cap--cylinder coexistence region
and the approach of both families to the common critical separatrix,
rather than an additional independent dynamical relation.

The geometric landmarks all occur at a few longitudinal pole lengths,
but their coefficients are shape dependent: the strip values are
approximately \(2.3\)--\(2.5\), whereas the spherical transition region
lies near \(3.6\). Such shape dependence is expected: the strip and
spherical area functionals weight the radial direction differently,
whereas \(\xi_\parallel^{\rm pole}\) is defined without reference to
any choice of boundary subregion. The comparison nevertheless
identifies a density scale absent from the unbackreacted closed-string
geometry: both the OSM landmarks and the pole-defined response length
scale as \(q^{-1/3}\). This shared scaling is fixed by dimensional
analysis, while their shape-dependent order-one coefficients show that
the agreement is parametric rather than a dynamical identity or a
universal numerical identification.
Appendix~\ref{app:longitudinal-poles} gives the derivation, higher
eigenvalues and numerical checks.

\section{Discussion}
\label{sec:discussion}

The open string metric remains a useful object. It governs probe-brane
fluctuations and captures effective horizons, effective temperatures and
crossover scales in the open-string sector. However, an RT-like minimal
surface in this metric, with the bare area functional, should not be
identified with the flavour entanglement entropy of the boundary theory.

Our finite-density D3--D7 example gives a sharp counterexample: in a
static state where the leading flavour entanglement entropy from
linearised D7-brane backreaction has shape-dependent volume terms, the
bare OSM areas instead saturate or show area-like behaviour in the same
controlled probe regime.

At the same time, the bare OSM area is sensitive to the density.
The leading closed-string RT surface in the unbackreacted probe geometry
does not see \(q\), while the OSM extremal surfaces do: the strip
saturation scale and the spherical branch endpoints are controlled by
\(q^{-1/3}\). The established probe-brane flavour EE contribution also
depends on \(q\), but it organises the large-region answer into
shape-dependent extensive terms rather than into the bare OSM scaling
found here \cite{ChangKarchUhlemann2014}.

The calculation in section~\ref{sec:static-longitudinal} quantifies
this relation rather than establishing an identity between the
geometric landmarks and a pole-defined response length. The nearest discrete
longitudinal pole lies on the imaginary axis, and the strip and sphere
landmarks occur at shape-dependent order-one multiples of its
pole-defined response length. The OSM surfaces therefore detect the
same density-induced open-sector scale, but not the shape-dependent
extensive flavour entanglement obtained from the leading linearised D7-brane
backreaction.

A related probe D3--D5 study found purely imaginary poles of the
Coulomb-dressed longitudinal propagator, implying monotonic exponential
charge screening without Friedel-like oscillations; that observable is
distinct from the pole of the ungauged global-current correlator studied
here \cite{AnantuaHartnollMartinRamirez2013}.

At finite temperature, this comparison can be extended to the
longitudinal probe dynamics. The finite-density D3--D7 system supports
zero sound at zero temperature \cite{KarchSonStarinets2009}. In the
low-temperature, high-density regime
\(\widehat q=q/(\pi T)^3\gg1\), its longitudinal response exhibits
collisionless-quantum, collisionless-thermal and hydrodynamic regimes
as functions of dimensionless frequency and momentum
\cite{DavisonStarinets2012}.

Within linear response, a natural dynamical comparison scale is
\(v_s\tau\), where \(v_s\) denotes the zero-sound velocity and \(\tau\)
the low-temperature probe-current relaxation time. Comparing this
length with the finite-temperature OSM branch scales would then test
whether the geometric diagnostic remains tied to the static density
scale or becomes sensitive to collective relaxation
\cite{ChenLucas2017}.

The infrared behaviour reported here is that of the bare area functional
continued to the singular OSM endpoint. The vanishing endpoint area
density and the \(\Lambda^{-5}\) suppression of its area tails show that
the result is not generated by a divergent deep-IR contribution.
Moreover, appendix~\ref{app:finite-temperature} embeds the calculation
in an equilibrium
finite-temperature family with a regular horizon. For strips, the
mismatch with the \(q\)-controlled flavour volume term can already be
stated in the controlled window
eq.~\eqref{eq:controlled_probe_window}, without taking a strict
deep-infrared limit. For spheres, the same appendix gives an analytic
OSM scaling argument rather than attributing a finite-temperature
spherical calculation to CKU\@. None of these observations turns the bare
OSM area into a replica-derived entropy.

From the viewpoint of driven probe-brane states, the present result
shows that the direct prescription explored in
ref.~\cite{BanerjeeBhattacharyaMaulik2021}, namely the unweighted OSM
area, cannot serve as a generally valid substitute for stationary
flavour EE\@. It cannot replace a
Lorentzian closed-string backreaction/HRT analysis merely because the
OSM correctly governs probe fluctuations and effective horizons.
Although the Chang--Karch framework can formally incorporate
worldvolume-gauge-field stresses and retarded bulk response
\cite{ChangKarch2013}, the complete stationary state additionally
depends on the mechanism that absorbs the injected energy and momentum.
The result of ref.~\cite{OBannonProbstRodgersUhlemann2017}, formulated
as an entanglement-rate law, constrains the time-dependent response in
its driving setup, not the complete EE of such a stationary state. Our
result does not exclude a more general weighted or replica-derived
functional involving open-string data; rather, it shows that such a
functional cannot be the bare OSM area alone.

A structural issue reinforces this conclusion. The OSM enters the
quadratic action for probe fluctuations together with a generally
running effective coupling \cite{KimShockTarrio2011}, and is induced
kinematically rather than obtained by extremising a
gravitational action \cite{BanerjeeBhattacharyaMaulik2021}. A canonical
Bekenstein--Hawking normalisation is therefore not supplied; this
obstruction was stated explicitly for the related D3--D7 worldvolume
horizon in ref.~\cite{OBannonProbstRodgersUhlemann2017}. Moreover, the
effective-horizon area need not track the effective temperature
\cite{NakamuraOoguri2013}, and entropy derived from the probe free
energy is not generally equal to the OSM horizon area
\cite{KunduKundu2013,BanerjeeKunduKundu2016}; see also
ref.~\cite{KunduReview2019}.

Different fluctuation sectors can moreover be described by effective
metrics that differ by position-dependent conformal factors, while
their quadratic actions retain channel-dependent prefactors or
effective couplings
\cite{KimShockTarrio2011,NakamuraOoguri2013,
BanerjeeBhattacharyaMaulik2021}. Since
codimension-two areas are not conformally invariant, the fluctuation
geometry alone cannot select a unique entropy functional. A complete
treatment of electrically driven stationary states would therefore
require a first-principles construction, such as a closed-string
backreaction/HRT analysis with an explicit prescription for the energy
sink, or a replica derivation of a more general functional involving
open-string data.

\acknowledgments

{\emergencystretch=1em
The author thanks Masataka Matsumoto and Norihiro Tanahashi for helpful
advice. \mbox{OpenAI's} ChatGPT and Codex, together with Anthropic's Claude,
were used for language editing, code assistance, and exploratory
analytical and numerical checks. All AI-assisted material was
independently verified by the author, who takes full responsibility for
the manuscript.\par}

\appendix

\section{Strip area comparison, subtraction and quadrature}
\label{app:strip-integrals}

This appendix records the strip formulas used in the numerical
comparison. Write
\begin{equation}
H(z)=1+q^2z^6 .
\end{equation}
On a constant-time slice, the strip area functional per transverse
volume is
\begin{equation}
\frac{{\cal A}_{\rm strip}}{V_2}
=
\int dx\,z^{-3}
\sqrt{1+\frac{z'(x)^2}{H(z)}}.
\end{equation}
Since the Lagrangian has no explicit \(x\)-dependence, the corresponding
first integral is
\begin{equation}
\frac{z^{-3}}{\sqrt{1+z'(x)^2/H(z)}}
=
\frac{1}{z_*^3},
\end{equation}
where \(z_*\) is the turning point. This gives
\begin{equation}
z'(x)^2
=
H(z)
\left(\frac{z_*^6}{z^6}-1\right),
\end{equation}
and hence the width formula in eq.~\eqref{eq:strip_width}.

The connected area is
\begin{equation}
\frac{{\cal A}_{\rm conn}}{V_2}
=
2\int_\epsilon^{z_*} dz\,
\frac{z_*^3}
{z^3\sqrt{H(z)}\sqrt{z_*^6-z^6}}.
\end{equation}
The disconnected reference surface consists of two vertical surfaces,
\begin{equation}
\frac{{\cal A}_{\rm disc}}{V_2}
=
2\int_\epsilon^\infty dz\,\frac{1}{z^3\sqrt{H(z)}}.
\end{equation}
Their difference is finite as \(\epsilon\to0\). Introducing
\(s=z/z_*\) and \(\zeta=q^{1/3}z_*\), one may write
\begin{equation}
\frac{z_*^2}{2V_2}
\left(
{\cal A}_{\rm conn}-{\cal A}_{\rm disc}
\right)
=
I_{\rm conn}(\zeta)-I_{\rm tail}(\zeta),
\label{eq:strip_area_difference}
\end{equation}
with
\begin{align}
I_{\rm conn}(\zeta)
&=
\int_0^1 ds\,
\frac{1}{s^3\sqrt{1+\zeta^6s^6}}
\left(
\frac{1}{\sqrt{1-s^6}}-1
\right),
\\
I_{\rm tail}(\zeta)
&=
\int_0^1 du\,
\frac{u^4}{\sqrt{u^6+\zeta^6}}.
\end{align}
The strip area-crossing point \(\ell_c\) is determined by the zero of
\({\cal A}_{\rm conn}-{\cal A}_{\rm disc}\). The maximal width
\(\ell_{\max}\) is instead the maximum of \(\ell(\zeta)\) along the
connected branch.

For stable numerical quadrature we used the change of variables
\(s^6=\sin^2 t\) in \(I_{\rm conn}\). This removes the apparent endpoint
singularity in the subtracted integral:
\begin{equation}
I_{\rm conn}(\zeta)
=
\frac{1}{3}\int_0^{\pi/2} dt\,
\frac{\sin^{-5/3}t\,\left(1-\cos t\right)}
{\sqrt{1+\zeta^6\sin^2t}}.
\end{equation}

\section{Spherical extremal-surface equations and infrared structure}
\label{app:spherical-eoms}

For spherical regions it is useful to keep the cap and cylinder
branches separate.

\subsection{Cap branch}

For \(z=z(\rho)\), the area functional is eq.~\eqref{eq:cap_area}. With
\(H(z)=1+q^2z^6\) and
\begin{equation}
S_{\rm cap}
=
\sqrt{1+\frac{z'(\rho)^2}{H(z)}} ,
\end{equation}
the Euler--Lagrange equation is
\begin{equation}
\frac{d}{d\rho}
\left(
\frac{\rho^2 z^{-3}z'}{H(z)S_{\rm cap}}
\right)
+
3\rho^2z^{-4}S_{\rm cap}
+
\frac{\rho^2z^{-3}z'^2 H'(z)}{2H(z)^2S_{\rm cap}}
=0.
\label{eq:cap_eom}
\end{equation}
Regularity at \(\rho=0\) fixes the local expansion
\begin{equation}
z(\rho)
=
z_*
-
\frac{H(z_*)}{2z_*}\rho^2
+
O(\rho^4),
\end{equation}
which is the shooting initial condition used to generate the cap branch.
In the numerical implementation we integrate the equivalent equation for
\(u(\rho)=z(\rho)^2\). This variable removes the square-root behaviour at
the AdS boundary and makes the extraction of the boundary radius more
stable.

\subsection{Cylinder branch}

For \(\rho=\rho(z)\), define
\begin{equation}
S_{\rm cyl}
=
\sqrt{\rho'(z)^2+\frac{1}{H(z)}}.
\end{equation}
The area functional in eq.~\eqref{eq:cylinder_area} gives
\begin{equation}
\frac{d}{dz}
\left(
\frac{\rho^2 z^{-3}\rho'}{S_{\rm cyl}}
\right)
-
2\rho z^{-3}S_{\rm cyl}
=0.
\label{eq:cylinder_eom}
\end{equation}
The large-\(z\) expansion of a cylinder branch is
\begin{equation}
\rho(z)
=
\rho_0
+
\frac{1}{10\rho_0 q^2 z^4}
-
\frac{1}{1000\rho_0^3q^4z^8}
+
O(z^{-10}).
\end{equation}
This is used as the infrared shooting condition. The existence of a
minimum boundary radius on this branch follows from the non-monotonic map
\(\rho_0\mapsto R\).

\subsection{Critical shrinking solution}

The finite-\(\rho_0\) expansion is non-uniform as \(\rho_0\to0\). A
separate shrinking asymptotic is obtained by setting
\(\rho(z)=Az^{-2}+\cdots\) in eq.~\eqref{eq:cylinder_eom}. The leading
residual is
\begin{equation}
\frac{2A\left(3A^2q^2-1\right)}
{q\sqrt{1+4A^2q^2}}\,z^{-8},
\end{equation}
so the non-trivial solution has \(A=1/(\sqrt{3}\,q)\). Extending the
ansatz to the next order gives
\begin{equation}
\rho_{\rm crit}(z)
=
\frac{1}{\sqrt{3}\,qz^2}
-
\frac{10}{59\sqrt{3}\,q^3z^8}
+O(z^{-14}).
\label{eq:critical_shrinking_full}
\end{equation}
Direct shooting of eq.~\eqref{eq:critical_shrinking_full} reproduces
\(R_\infty q^{1/3}=0.9917717\), in agreement with the limiting values
obtained from both ordinary branch families.

More generally, a dominant-balance analysis of
\(\rho\sim Az^p\) leaves only the finite-cylinder behaviour \(p=0\) and
the critical shrinking behaviour \(p=-2\). The latter is the unique
separatrix approached by the limits \(\rho_0\to0\) and
\(z_*\to\infty\).

Linearising around a finite-\(\rho_0\) cylinder also produces a formal
growing mode. This mode does not define an additional nonlinear
asymptotic branch: direct substitution of a conical ansatz
\(\rho\sim\alpha z\) into eq.~\eqref{eq:cylinder_eom} leaves a
non-vanishing leading residual proportional to
\(-3\alpha^2/z^2\). Thus no asymptotically conical power-law branch
supplements the cap, finite-cylinder and critical solutions described
above.

\section{Numerical implementation and systematic checks}
\label{app:numerics}

The numerical work uses dimensionless variables. Equivalently, we set
\(q=1\) in the code and report the corresponding dimensionless
combinations. For general \(q>0\), a code value \(\widehat L\)
corresponds to the physical length \(L=q^{-1/3}\widehat L\). Thus the
reported quantities are
\[
q^{1/3}\ell,\qquad q^{1/3}R,\qquad q^{1/3}z_*,
\qquad q^{1/3}\rho_0 .
\]
Python supplied the principal numerical results, while Mathematica
provided an independent cross-check.

For strips, the width integral and the finite area difference
\eqref{eq:strip_area_difference} are one-dimensional integrals. The
strip area-crossing point is the root of the finite area difference, and
the maximum width is found by a one-dimensional maximisation of
\(\ell(\zeta)\).

For spherical caps, the shooting variable is \(u=z^2\). Starting from
the regular expansion near \(\rho=0\), the integration is stopped at a
small \(u_\epsilon\). The boundary radius is obtained by linear
extrapolation in \(u\),
\begin{equation}
R
=
\rho_{\rm end}
-
\frac{u_{\rm end}}{u'_{\rm end}}.
\end{equation}
Here the subscript `end' denotes the final integration point.
For cylinders, we integrate in \(x=\log z\), which avoids a large
hierarchy between the infrared starting point and the AdS boundary. Near
the boundary,
\begin{equation}
\rho(z)
=
R-\frac{z^2}{2R}+O(z^4),
\end{equation}
so, with \(v=d\rho/dx=z\,d\rho/dz\), the numerical boundary radius is
extracted as
\begin{equation}
R
=
\rho(z_\epsilon)-\frac{1}{2}v(z_\epsilon).
\end{equation}
For the sphere area comparison in the overlap window, both cap- and
cylinder-type surfaces are integrated in the \(\rho(z)\) representation
down to a common cutoff \(z_\epsilon\). We subtract
\(2\pi R^2/z_\epsilon^2+2\pi\log z_\epsilon\) from each area. The
subtraction cancels in
\({\cal A}_{\rm cap}^{\ren}(R)-{\cal A}_{\rm cyl}^{\ren}(R)\) at fixed
\(R\).

Because these are deterministic calculations, we do not interpret
numerical variation as a statistical sampling error. For the sphere
crossing we instead estimate a numerical-systematics envelope. We
repeated the full shooting and fixed-\(R\) area comparison at
\(z_\epsilon=2.0\times10^{-3},10^{-3},7.5\times10^{-4}\) and rescaled
the ODE solver tolerances by factors \(2\) and \(0.4\). The resulting
values lie in the range
\begin{equation}
0.9897087
\le R_\times q^{1/3}\le
0.9897760.
\end{equation}
The independent Mathematica result, \(0.9897102\), lies within this
range. We therefore round the envelope upward and quote
\(R_\times q^{1/3}=0.9897(1)\), where the parenthetical \(1\) denotes an
absolute numerical uncertainty of \(10^{-4}\); it is not a statistical
confidence interval. The root-finder tolerance, \(8\times10^{-7}\), is
smaller than this systematic envelope.

\section{Finite-temperature regularisation of the OSM endpoint}
\label{app:finite-temperature}

It is useful to check that the zero-temperature conclusion is not an
artefact of continuing the bare OSM area functional to the singular
\(z\to\infty\) endpoint. At small but finite temperature, define
\begin{equation}
f(z)=1-\left(\frac{z}{z_h}\right)^4,
\qquad
H(z)=1+q^2z^6 .
\end{equation}
No external electric field is present. This is therefore an equilibrium
regulator rather than an electric-field-driven NESS construction.
The finite-density radial worldvolume field strength \(F_{zt}=A_t'\) is
unchanged,
\begin{equation}
A_t'(z)=\frac{qz}{\sqrt{H(z)}}.
\end{equation}
For the massless embedding this follows because
\(g_{tt}g_{zz}=-z^{-4}\) is independent of \(f(z)\); the massless
embedding itself remains an exact solution at finite temperature.
The five-dimensional finite-temperature OSM sector is
\begin{equation}
ds_{\OSM}^2
=
\frac{1}{z^2}
\left[
-\frac{f(z)}{H(z)}dt^2
+d\vec x^{\,2}
+\frac{dz^2}{f(z)H(z)}
\right].
\end{equation}
The OSM horizon is at \(z=z_h\). Since the factor \(H(z_h)\) appears
symmetrically in \(G_{tt}\) and \(G_{zz}\), the OSM temperature is the
background temperature,
\begin{equation}
T_{\OSM}=T=\frac{1}{\pi z_h}.
\end{equation}
The horizon spatial area density is
\begin{equation}
\sqrt{G_{xx}G_{yy}G_{ww}}\big|_{z=z_h}
=
z_h^{-3}
=
(\pi T)^3 .
\end{equation}
This result is exact at arbitrary \(q\) within the present setup:
\(G_{ij}=\delta_{ij}/z^2\) contains no factor of \(H(z)\), so no mixed
density contribution can enter the horizon volume coefficient.
Thus finite-temperature OSM minimal areas can have a thermal volume
coefficient \(z_h^{-3}\), but this coefficient vanishes as \(T\to0\).
The regulated family therefore has no \(q\)-controlled thermal volume
coefficient surviving as \(T\to0\). This is consistent with the
zero-temperature saturation and \(R^2\) scaling: the regulator does not
restore the \(q\)-controlled volume coefficient of the flavour
benchmark. The strip comparison and the corresponding parametric OSM
statement for spheres are separated below.

For a strip, the finite-temperature connected branch obeys
\begin{equation}
\frac{\ell}{2}
=
\int_0^{z_*}
\frac{z^3\,dz}
{\sqrt{f(z)H(z)}\sqrt{z_*^6-z^6}},
\end{equation}
and
\begin{equation}
\frac{{\cal A}_{\rm conn}}{V_2}
=
2\int_\epsilon^{z_*}
\frac{z_*^3\,dz}
{z^3\sqrt{f(z)H(z)}\sqrt{z_*^6-z^6}}.
\end{equation}
The corresponding disconnected reference area is
\begin{equation}
\frac{{\cal A}_{\rm disc}}{V_2}
=
2\int_\epsilon^{z_h}
\frac{dz}{z^3\sqrt{f(z)H(z)}}
+
\frac{\ell}{z_h^3}.
\end{equation}
For \(\vartheta\equiv\pi Tq^{-1/3}\ll1\), the zero-temperature endpoint
tails removed by cutting the geometry off at \(z_h\) obey
\begin{equation}
\frac{\Delta {\cal A}_{\rm strip}^{(0)}}{V_2}
\sim \frac{2}{5qz_h^5},
\qquad
\frac{\Delta {\cal A}_{\rm cyl}^{(0)}}{4\pi R^2}
\sim \frac{1}{5qz_h^5}.
\label{eq:finite-t-endpoint-tails}
\end{equation}
The estimates in eq.~\eqref{eq:finite-t-endpoint-tails} refer only to
the endpoint tails removed by the finite-temperature cutoff. They are
not the complete finite-temperature correction: over the density-scale
region \(z\sim q^{-1/3}\), expanding \(f(z)^{-1/2}\) gives relative
corrections of order \(\vartheta^4\), which are still parametrically
small.
The cylinder estimate is the large-\(R\) spherical result. Both
normalised tails are of order \(q^{2/3}\vartheta^5\). Matching the
zero-temperature finite area scale \(q^{2/3}\) to the thermal volume
coefficient \(z_h^{-3}\) gives, for a generic region size \(L\), the
crossover scale
\begin{equation}
L_T\sim q^{2/3}z_h^3
=q^{-1/3}\vartheta^{-3}.
\label{eq:thermal-crossover-scale}
\end{equation}
Here \(L=\ell\) for strips and \(L=R\) for spheres. For strips, one
therefore expects a parametrically broad quasi-plateau for
\(q^{-1/3}\ll\ell\ll L_T\), followed at still larger widths by weak
thermal linear growth; its slope vanishes as \(T\to0\).
For \(\vartheta\ll1\), the CKU finite-temperature analysis shows that
the strip volume term persists in the controlled window
\(q^{-1/3}\ll\ell\ll z_h\)
\cite{ChangKarchUhlemann2014}. In the same window, the bare OSM strip
diagnostic remains parametrically close to its zero-temperature
saturation; its thermal volume term becomes competitive only at
\(L_T\gg z_h\). The strip mismatch can therefore be formulated without
taking a strict deep-infrared limit.

For spheres, the CKU benchmark used here is instead the
zero-temperature large but finite probe result. The analogous hierarchy
\(q^{-1/3}\ll R\ll z_h\) is a parametric expectation, not an explicit
finite-temperature CKU calculation. Independently, the OSM tail and
scale estimates above imply that its spherical area remains close to
the zero-temperature \(R^2\) behaviour throughout that hierarchy. At
still larger sizes, it is governed by the horizon area density and has
a thermal volume term with coefficient \(z_h^{-3}\). Detailed
finite-temperature swallowtail numerics are not needed for this OSM
scaling comparison.

\section{Probe-brane flavour entanglement: backreaction and replica
methods}
\label{app:probe-flavour-ee}

This appendix summarises the two complementary leading-order methods
for computing probe-brane contributions to boundary entanglement
entropy and clarifies which method underlies the finite-density
benchmark used in section~\ref{sec:benchmark}. Let \(t_0\) denote the
dimensionless probe parameter, proportional to \(N_f/N_c\) at fixed
't Hooft coupling. For a boundary region \(A\), the entropy has the
expansion
\begin{equation}
S_A
=
S_A^{(0)}
+
S_{\fl}^{(1)}(A)
+
O\!\left(t_0^2 S_A^{(0)}\right),
\qquad
\frac{S_{\fl}^{(1)}(A)}{S_A^{(0)}}=O(t_0).
\label{eq:probe-expansion}
\end{equation}
Here \(S_A^{(0)}\) is the leading adjoint-sector entropy and
\(S_{\fl}^{(1)}(A)\) is its first probe-flavour correction.
There is no contradiction between the probe limit and the appearance of
a linearised backreaction in \(S_{\fl}^{(1)}(A)\): the latter is
precisely the first correction in the probe expansion and already
carries its physical probe normalisation.

\subsection{Linearised-backreaction route}
\label{app:linearised-backreaction}

The conceptually direct procedure is to compute the metric perturbation
\(h_{\mu\nu}\) sourced by the probe-brane stress tensor
\(T_{\mu\nu}^{\rm probe}\),
\begin{equation}
\mathcal E_{\mu\nu}{}^{\rho\sigma}h_{\rho\sigma}
=
8\pi G_N\,T_{\mu\nu}^{\rm probe},
\label{eq:linearised-einstein-probe}
\end{equation}
where \(G_N\) is the bulk Newton constant and
\(\mathcal E_{\mu\nu}{}^{\rho\sigma}\) is the linearised Einstein
operator. One then evaluates the first variation of the closed-string
RT area. If
\(\gamma_A^{(0)}\) is the extremal surface in the unperturbed geometry,
its first-order embedding variation vanishes by extremality, and the
entropy correction is
\begin{equation}
S_{\fl}^{(1)}(A)
=
\frac{1}{8G_N}
\int_{\gamma_A^{(0)}} d^{d-1}\sigma\,
\sqrt{\gamma^{(0)}}\,
\gamma_{(0)}^{ab}\,
h_{\mu\nu}(X)
\partial_aX^\mu\partial_bX^\nu .
\label{eq:area-linear-response}
\end{equation}
Here \(d\) is the boundary spacetime dimension,
\(\gamma_{ab}^{(0)}\) is the induced metric on
\(\gamma_A^{(0)}\), and \(X^\mu(\sigma)\) is its embedding.
Chang and Karch express \(h_{\mu\nu}\) through the linearised
gravitational Green function and thereby rewrite
eq.~\eqref{eq:area-linear-response} as a compact double integral
involving the probe stress tensor and the unperturbed RT surface
\cite{ChangKarch2013}. Their method systematically includes the leading
backreaction, but does not require construction of the complete
backreacted geometry point by point or a new solution for the extremal
surface in that geometry.

Worldvolume gauge fields enter this construction through
\(T_{\mu\nu}^{\rm probe}\). More generally, if the probe sources
additional bulk fields that mix linearly with fields present in the
background, their induced metric response must also be included. For
time-dependent perturbations, the corresponding retarded bulk Green
function is required. Independently of the Chang--Karch reformulation,
ref.~\cite{KontoudiPolicastro2014} computed the entanglement entropy
directly using a known first-order backreacted massive D3--D7 geometry.
This is a concrete backreacted-geometry calculation rather than the
compact Green-function construction above. The finite-density
D3--D7 result of CKU used as the benchmark in
section~\ref{sec:benchmark} employs the Chang--Karch
linearised-backreaction route
\cite{ChangKarch2013,ChangKarchUhlemann2014}.

\subsection{Euclidean probe-action route}

Building on generalised gravitational entropy
\cite{LewkowyczMaldacena2013}, Karch and Uhlemann adapted the replica
construction to probe branes \cite{KarchUhlemann2014}. We choose the
convention in which the Euclidean angular coordinate around the
entangling surface has period \(2\pi n\), and denote the corresponding
replicated bulk saddle by \(B_n\). Suppressing inessential notation, the
leading probe contribution has the schematic
structure\footnote{Writing
\(I_n=I_{\rm brane}^{\rm ren}[B_n]\), the full-period convention gives
\(\left.(n\partial_n-1)I_n\right|_{n=1}\). Karch and Uhlemann instead
use an action \(\widetilde I_n\) with a fixed \(2\pi\) angular
integration range, for which the expression is
\(\left.n^2\partial_n\widetilde I_n\right|_{n=1}\). Since
\(I_n=n\widetilde I_n\), the two formulae are equivalent.}
\begin{equation}
S_{\fl}^{(1)}(A)
=
\left.
\left(n\partial_n-1\right)
I_{\rm brane}^{\rm ren}[B_n]
\right|_{n=1}.
\label{eq:probe-replica}
\end{equation}
Here \(n\) is the replica index and
\(I_{\rm brane}^{\rm ren}[B_n]\) is the renormalised Euclidean on-shell
probe-brane action evaluated on \(B_n\). It includes the physical
probe-brane normalisation, holographic counterterms and any required
worldvolume boundary terms. No additional factor of \(t_0\) multiplies
eq.~\eqref{eq:probe-replica}; its probe-order normalisation is already
contained in the brane action.

In overlapping static examples, the Euclidean probe-action method
reproduces the leading result obtained from linearised backreaction,
subject to the usual counterterm and boundary-term qualifications
\cite{ChangKarch2013,KarchUhlemann2014}. A recent comparison with fully
backreacted flavoured geometries found agreement in the examples
studied after omitting a specific infrared boundary term arising from a
total derivative; the status of this term remains subtle
\cite{JokelaEtAl2024}. For spherical regions centred on a
conformal defect, Jensen and O'Bannon developed a complementary
hyperbolic-space construction. It agrees with the leading-backreaction
result in overlapping probe examples
\cite{JensenOBannon2013,ChangKarch2013}.

\subsection{Scope of the present benchmark}

The finite-density CKU result used here follows the linearised
backreaction method; it was not obtained by applying
eq.~\eqref{eq:probe-replica} directly. Its zero-temperature large-region
expansion is controlled only for large but finite regions, before the
probe backreaction becomes non-perturbative in the deep infrared. For
strips, CKU's finite-temperature analysis identifies the controlled
window in eq.~\eqref{eq:controlled_probe_window}. For spheres, the
benchmark invoked here is instead the zero-temperature large but finite
probe regime; the analogous finite-temperature hierarchy is only a
parametric expectation.

The bare OSM area studied in the main text is not obtained by either of
the methods above. The OSM is an effective metric governing worldvolume
fluctuations, but in the present setup it is not a dynamical
gravitational field with a known replica saddle, Newton constant or
homology prescription. The purpose of this paper is precisely to test
whether its unweighted codimension-two area reproduces the established
leading flavour EE\@. The comparison shows that it does not in the
finite-density state considered here.

For driven states, leading backreaction can include electric fields and
retarded bulk response. No direct Euclidean implementation of
eq.~\eqref{eq:probe-replica} is established for \(A_x=-Et\), and a NESS
also requires specifying the energy and momentum sink.
Ref.~\cite{OBannonProbstRodgersUhlemann2017} gives an entanglement-rate
law, not the complete stationary-state EE\@.

\section{Static longitudinal response poles}
\label{app:longitudinal-poles}

This appendix derives the longitudinal response calculation used in
section~\ref{sec:static-longitudinal}. We restrict attention to the
longitudinal channel.\footnote{The transverse static
equation has \(a_\perp=A+B/u+\cdots\) at large \(u\), while its
Sturm--Liouville weight behaves as
\(w_\perp=1/[u\sqrt{1+u^6}]\sim u^{-4}\). Both branches therefore have
finite norm, so normalisability alone does not determine a unique
infrared boundary condition; we do not use this channel below. By
contrast, the longitudinal weight behaves as \(u^2\), and
normalisability uniquely selects the decaying \(u^{-7}\) branch.}
Following
ref.~\cite{KarchSonStarinets2009}, write the longitudinal worldvolume
gauge-field fluctuations as \(a_t(z)\) and \(a_x(z)\), with frequency
\(\omega\), momentum \(k\) and Fourier dependence
\(e^{-i\omega t+ikx}\). Their gauge-invariant combination is
\(\mathcal E=\omega a_x+k a_t\). For the massless D3--D7 system at
zero temperature, the longitudinal equation at \(\omega=0\) reduces to
\begin{equation}
\mathcal E''(z)
+
\left(
\frac{3H'(z)}{2H(z)}-\frac{1}{z}
\right)
\mathcal E'(z)
-
\frac{k^2}{H(z)}
\mathcal E(z)
=0,
\qquad
H(z)=1+q^2z^6 .
\label{eq:longitudinal-dimensional}
\end{equation}
With \(k=i\kappa_\parallel\), \(u=q^{1/3}z\) and
\(c_\parallel=\kappa_\parallel/q^{1/3}\), this is
eq.~\eqref{eq:longitudinal-static-pole}. Equivalently, it is the
Sturm--Liouville problem
\begin{equation}
-\frac{d}{du}
\left[
\frac{(1+u^6)^{3/2}}{u}
\frac{d\mathcal E}{du}
\right]
=
c_\parallel^2
\frac{\sqrt{1+u^6}}{u}
\mathcal E .
\label{eq:longitudinal-sturm-liouville}
\end{equation}
For functions obeying the source-free ultraviolet condition and the
normalisable infrared condition, the boundary term in the
Sturm--Liouville identity vanishes. Hence
\begin{equation}
c_\parallel^2
=
\frac{
\displaystyle\int_0^\infty du\,
\frac{(1+u^6)^{3/2}}{u}
\left|\mathcal E'(u)\right|^2
}{
\displaystyle\int_0^\infty du\,
\frac{\sqrt{1+u^6}}{u}
\left|\mathcal E(u)\right|^2
}
>0 .
\label{eq:longitudinal-rayleigh}
\end{equation}
The discrete static poles selected by these boundary conditions
therefore occur in pairs
\(k=\pm i c_\parallel q^{1/3}\) on the imaginary \(k\)-axis.

Near the AdS boundary the two independent behaviours may be written as
\begin{equation}
\mathcal E(u)
=
A\bigl[1+O(u^2\log u)\bigr]
+
B u^2
\left[
1-\frac{c_\parallel^2}{8}u^2
+\frac{c_\parallel^4}{192}u^4
+O(u^6)
\right].
\label{eq:longitudinal-uv}
\end{equation}
The coefficient \(A\) is the boundary source, so the pole condition is
\(A=0\). At the zero-temperature infrared endpoint, the normalisable
solution is
\begin{equation}
\mathcal E(u)
=
u^{-7}
\left[
1-\frac{c_\parallel^2}{44u^4}
-\frac{21}{26u^6}
+O(u^{-8})
\right],
\qquad
u\longrightarrow\infty .
\label{eq:longitudinal-ir}
\end{equation}
This condition uniquely fixes the longitudinal endpoint behaviour.

For the numerical shooting, we integrate the endpoint-adapted fields
\(\phi=\mathcal E/u^2\) from the ultraviolet and
\(\psi=u^7\mathcal E\) from the infrared. Their Wronskian is set to zero
at \(u=1\), and each root is isolated by a sign-changing bracket. The
first three eigenvalues are
\begin{equation}
c_{\parallel,1}=3.6334359,
\qquad
c_{\parallel,2}=6.1094883,
\qquad
c_{\parallel,3}=8.4444131 .
\label{eq:longitudinal-first-three}
\end{equation}
Changing the integration endpoints and solver tolerances leaves all
displayed digits unchanged.

The low-temperature regulator gives a complementary endpoint check.
Writing
\begin{equation}
f_T(u)=1-\left(\frac{u}{u_h}\right)^4,
\qquad
u_h=q^{1/3}z_h,
\label{eq:longitudinal-finite-t-definitions}
\end{equation}
the static equation becomes
\begin{equation}
\frac{d}{du}
\left[
\frac{(1+u^6)^{3/2}}{u}
\frac{d\mathcal E}{du}
\right]
+
c_\parallel^2
\frac{\sqrt{1+u^6}}{u f_T(u)}
\mathcal E
=0 .
\label{eq:longitudinal-finite-t}
\end{equation}
Regularity of the static worldvolume potential imposes
\(\mathcal E(u_h)=0\). Table~\ref{tab:longitudinal-regulator-convergence}
shows the convergence of the lowest regulated eigenvalue as \(u_h\) is
increased.
\begin{table}[ht]
\centering
\begin{tabular}{cc}
\toprule
\(u_h\) & \(c_{\parallel,1}\) \\
\midrule
\(5\)  & \(3.6325615\) \\
\(8\)  & \(3.6333005\) \\
\(12\) & \(3.6334090\) \\
\(20\) & \(3.6334324\) \\
\(30\) & \(3.6334352\) \\
\bottomrule
\end{tabular}
\caption{
Convergence of the lowest longitudinal eigenvalue
\(c_{\parallel,1}\) as the regulated horizon position \(u_h\) is
increased.
}
\label{tab:longitudinal-regulator-convergence}
\end{table}

The sequence converges to the zero-temperature value in
eq.~\eqref{eq:longitudinal-first-three}, confirming that the decaying
longitudinal branch is selected by the regular finite-temperature
family.

\FloatBarrier

\bibliographystyle{JHEP}
\bibliography{references}

\end{document}